\documentclass[9pt]{article}
\usepackage{latexsym}
\usepackage{graphicx}
\usepackage{amsfonts,amssymb}
\begin{document}
\title{\textbf{Community Detecting By Signaling on Complex Networks}}
\author{Yanqing Hu, Menghui Li, Peng Zhang, Ying Fan, Zengru Di\footnote{Author for correspondence: zdi@bnu.edu.cn}\\
\\\emph{Department of Systems Science, School of Management}\\
\emph{Center for Complexity Research}\\
\emph{Beijing Normal University, Beijing 100875, P.R.China}}
\maketitle
\begin{abstract}
Based on signaling process on complex networks, a method for
identification community structure is proposed. For a network with
$n$ nodes, every node is assumed to be a system which can send,
receive, and record signals. Each node is taken as the initial
signal source once to inspire the whole network by exciting its
neighbors and then the source node is endowed a $n$d vector which
recording the effects of signaling process. So by this process, the
topological relationship of nodes on networks could be transferred
into the geometrical structure of vectors in $n$d Euclidian space.
Then the best partition of groups is determined by $F$-statistic and
the final community structure is given by Fuzzy $C$-means clustering
method (FCM). This method can detect community structure both in
unweighted and weighted networks without any extra parameters. It
has been applied to ad hoc networks and some real networks including
Zachary Karate Club network and football team network. The results
are compared with that of other approaches and the evidence
indicates that the algorithm based on signaling process is
effective.
\end{abstract}



\section{Introduction}
The study of complex networks has received an enormous amount of
attention\cite{1,Newman reviwe,new reviwe} from the scientific
community in recent years. Physicists in particular have become
interested in the study of networks describing the topologies of
wide variety of systems, such as the world wide web\cite{www},
social and communication networks\cite{social network1,social
network2}, biochemical networks and many more\cite{metabolism}. One
such problem is the analysis of community structure found in many
networks. Distinct communities or modules within networks can
loosely be defined as subsets of nodes which are more densely
linked, when compared to the rest of the network\cite{14,15}. Such
communities have been observed, using some of the methods, in many
different contexts, including metabolic networks, banking networks
and most notably social networks. As a result, the problem of
identification of communities has been the focus of many recent
efforts. Community detection in large networks is potentially very
useful. Nodes belonging to a tight-knit community are more than
likely to have other properties in common. Besides, these
communities may probably be functional groups, which provide us
valuable reference to our study in many other fields. In recent
studies, the scientists have designed many different
algorithms\cite{14,15,2,3,4,5,6,7,8,9,10,11,12,13,18,19,20}(see
\cite{4} as a review) to detect the community structures. These
algorithms can be divided into categories. Some algorithms are
designed according to maximal modularity $Q$. Some are designed
based on topology structures (betweenness, degree, or clustering
coefficient). The last is designed according to the dynamical
properties of the network.

Communities within networks can be defined as subsets of nodes which
are more densely linked, when compared to the rest of the network.
Modularity $Q$ is an index advanced by Newman and Grivan\cite{22} as
a measurement for the community structure. It gives a clear and
precise definition of characteristics of the acknowledged community
and have very successful application in practice. So it leads to
many other algorithms brought forward to divide a community by
maximizing Modularity $Q$. Unfortunately, maximizing Modularity $Q$
has been proven a NPC problem\cite{Q NPC}, which makes it unable to
work out the partition corresponding to maximal $Q$ in a network
which has lots nodes. Actually many algorithms for maximizing $Q$
are usually heuristic algorithm. Besides, Modularity $Q$ has been
proven strictly that, as an index to measure the community
structure, it tends to combine the little communities rather than
identify them successfully in the networks with definite
communities\cite{23}. Though Modularity $Q$ is proven to have the
above-mentioned inherent defects, it is still the successful index
to measure a network for the moment. Therefore, lots of work for
detecting communities are dependent on the index $Q$.

Among the algorithms based on network topology, we want to introduce
briefly the spectral analysis method and GN algorithm\cite{15}. GN
algorithm was proposed by Grivan and Newman. It gives the division
of network by remove links. The links with largest betweenness are
removed one by one in order to split hierarchically the graph into
community. But GN algorithm alone can only give the dendrogram
concerning the network structure finally. It could not give the best
partition directly. If we want to get the best partition we have to
depend on Modularity $Q$ or other indexes to work it out. While the
principle of the spectral analysis method\cite{17} is based on the
theory of eigenvector of matrix. When a network is partitioned into
two communities with the pre-fixed sizes, the best partition is to
make the number of edges between the two communities being minimum.
In fact, this problem is closely related to the eigenvector
corresponding with the Fiedler eigenvector (the eigenvector
correspond to second minimum eigenvalue) of Laplacian matrix of a
network. Relatively speaking, spectral analysis method is the most
math theory-based approach. But it's still unable to get the best
partition. Besides, it requires to know the sizes of the two
sub-network beforehand. Although some methods are proposed to solve
the two defects, like getting the best partition by using $Q$,
ascertaining the sizes of the two sub-network by the sign of the
elements of Fielder eigenvector and so on, the inherent defects of
spectral analysis method have not been solved perfectly and
completely yet.

There are still other algorithms based on the dynamics on networks,
among which random walks\cite{6} method and circuit approach
method\cite{9} will be briefly discussed here. In the random walks
method, each node contains a walker initially. Then each walker will
randomly choose a neighbor of the node it currently stand on to
localize. This is a Markov process. After a period of time, the
possibility will be higher that the walker reach another node
belonging to the same community of the node the walker stood on at
the beginning. And this possibility can be directly regarded as the
possibility of pair of nodes in the same community and through it, a
dendrogram can be got, then partition can be made by the aid of
Modularity $Q$.

When using random walk method to detect communities, it's difficult
to specify the optimum random-walking time. And the best partition
is dependent on the external index $Q$ usually. The principle of
circuit approach method is to regard the edges of the network as the
resistances, and add voltage to the adequate nodes of the network,
then work out the voltage of each node by Kirchhoff's law. Nodes
with the similar voltages are regarded to exist in the same
community more probably. At the same time, define the external
indexes such as tolerance to realize the partition of the network.

In this paper, we propose a new algorithm for identification
communities based on signaling process on network. In this approach,
every node is viewed as a system which can be inspired. It can send,
receive, and record signals. In the initial, a node is selected as
the source of signal. We give it an initial unit signal and other
nodes with the signal of zero. Then the source node send the signal
to its neighbors and itself first. Afterwards, the nodes with
signals can also send signals to their neighbors and themselves. One
thing should be mentioned in this signaling process is that the node
can record the amount of signals it received, and at every time
step, each node sends its present-owning signals to its neighbors
and itself. After the inspiration of a certain $T$ time steps, the
signal distribution over the nodes could be viewed as the influence
of the source node to the whole network. For a network with $n$
nodes, signal distribution can be characterized by a $n$d vector.

If a network has $n$ nodes, we can get the influence of every node
by the same operation. The results are given by $n$ $n$d vectors.
Generally speaking, the source node should influence its community
first then through the community to influence the whole network. So
naturally, compared with the nodes in other communities, nodes of
the same community have similar influence toward the whole network.
And the difference of influence could be given by the $n$d vectors.

Thus, by the above signaling process on networks, the topological
structure of nodes is converted into the geometrical relationships
of vectors in $n$d Euclidian space. We can get the community
structures of nodes by clustering these $n$d vectors. Actually,
there are already a lots of methods to cluster vectors in Euclidian
space. Here we chose Fuzzy C-means clustering method (FCM) assisted
by $F$ statistic\cite{24} to get the best partition of the
communities. $F$ statistic is developed in mathematical statistics.
It describes the best partition as the one with the shorter average
distance between the vectors inside the same community, and the
larger distance between vectors of different communities. After
getting the best number of groups by $F$ statistic, we can work out
the communities by FCM. With the aid of $F$ statistic, the method
presented in this paper can detect community structures in complex
network without any extra parameters.

Some problems related to above method are also discussed, including
the optimum time steps $T$ of inspiration and the generalization of
the method to weighted networks. Then we applied the method to
detect the communities in \emph{ad hoc} and some real networks. Its
precision and accuracy are obtained and compared with some other
algorithms. The results indicate that the method based on signaling
process performs good.

\section{Method Based on Signaling Process}
\subsection{Basic Algorithm}
\textbf{A. Signaling Process. } For a network with $n$ nodes, every
node is assumed to be a system which can send, receive, and record
signals. One node can only affect its neighbors which will affect
their neighbors too in the same way. Finally, each node will affect
the whole network. In general, one node will affect its community
first and then the whole network via its community. So we can safely
conclude that the nodes in the same community will affect the whole
network in a similar way.

\begin{figure}
\center
\includegraphics[width=14cm]{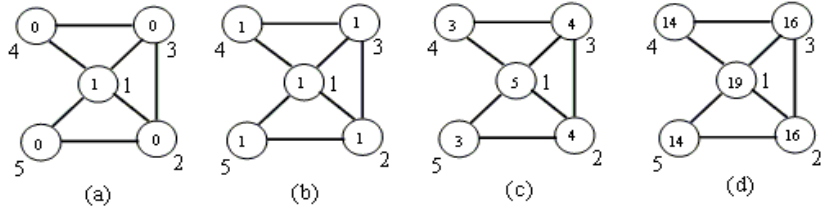}
\caption{Sketch of signaling process.(a)We set a signal on node $1$
and other nodes have no signal initially. (b)In the first step node
$1$ sends one signal to all its neighbors which are nodes $2,3,4$
and $5$ and itself. Then all nodes have $1$ signal. (c)Next, each of
them will send one signal to their neighbors and themselves
respectively at the same time. After the second step, node $1$ has
$5$ signals, both node $2$ and $3$ have $4$ signals and node $4$ and
$5$ have $3$ signals. Vector $[5,4,4,3,3]$ represents the effect of
the node $1$ to the whole network in $2$ steps. (d)Then every nodes
send the same amount of signals as they received in the last step to
their neighbors and themselves.}\label{spread in a simple net}
\end{figure}

At the beginning, we select a node as source and let it has one unit
of signal and the other nodes have no signal. Then let the source
node send signal to all of its neighbors and itself. After the first
step the node and all its neighbors have a signal. In the second
step, all the nodes which have signal will send it to their
neighbors and themselves. Every node will record the amount of
signals it received and then it will send the same quantity of
signals in the next time step. In this way, the process will be
repeated constantly on the network. After $T$ time steps, we can get
a $n$d vector that records each node's signal quantity which
represents the effect of the source node. The signaling process is
sketched out in Fig.\ref{spread in a simple net} by simple network
with 5 nodes. Choosing every node as the source node respectively,
we can get $n$ such vectors. The purpose that we let each node sends
signal or signals to itself is to take account of the historical
effects. This has been proved to be helpful to distinguish the
amounts of signals between the nodes in the community and outside in
a relatively short time period. Standardizing the $n$ vectors, then
the distance of each pair of vectors will represent the similarity
of the corresponding nodes. Using this kind of similarity the
communities can be detected.

Actually, the above signaling process could be described by a simple
but clear mathematical mechanism. Suppose we have a network with $n$
nodes, it can be represented mathematically by an adjacency matrix
$\textbf{A}$ with elements $A_{ij}$ if there is an edge from $i$ to
$j$ and $0$ otherwise. Then the column $i$ of matrix
$\textbf{V}=(\textbf{I}+\textbf{A})^{T}$ will represent the effect
of source node $i$ to the whole network in $T$ steps. In order to
get the relative effect, we should standardize each column of matrix
$\textbf{V}$. Assume the column $i$ of $\textbf{V}$ is
$V_{i}=(v_{i1},v_{i2}...v_{in})$, then the $V_{i}$ can be
standardized as $U_{i}=(u_{i1},u_{i2}...u_{in})$, here
$u_{ij}=\frac{v_{ij}}{\sum_{j}^{n}v_{ij}}$. Then to partition the
network which has $n$ nodes is equivalent to cluster $n$ vectors
$U_{1},U_{2},\cdots,U_{n}$ in Euclidian space.

\textbf{B. Fuzzy $C$-mean clustering. }It is well known that there
are many clustering methods and algorithms for the vectors in
Euclidian space. In this paper, we choose the inexpensive fuzzy
$C$-mean clustering algorithm (FCM)\cite{24} to detect communities
for the vectors given by signaling process. FCM is described as
following.
\begin{enumerate}
\item  Set $C$ as the number of communities to
partition.

\item  Randomly choose $C$ vectors for the $C$ communities as their
barycenters.

\item  Randomly choose a vector. The vector will belong to the
community when the distance between the vector and the barycenter of
the community is minimum among all the barycenter of communities.

\item  Re-compute the communities' barycenters which have added a
vector or deleted a vector.

\item  Repeat step 3 to step 4 until all the barycenters cannot be
modified.
\end{enumerate}

We know that there are many definitions of distance. In our
algorithm we choose the normal definition--Euclidian distance to
measure the similarity between vectors of nodes.

\textbf{C. $F$ Statistic.} At the first step of fuzzy $C$-mean
clustering algorithm, we must set an extra parameter $C$ which
presents how many clusters we will partition. Here we use $F$
statistic\cite{24} to estimate the proper $C$. Now let we have a
glance at $F$ statistic. Suppose $U=\{u_{1},u_{2},\cdots,u_{n}\}$ is
the set of vectors of all nodes and
$u_{j}=(x_{j1},x_{j2},\cdots,x_{jn})$, here $x_{jk}$ is the $k$th
character quantity of $u_{j}$. Suppose $C$ is the number of
communities and $n_{i}$ is the number of nodes of $i$th community.
We name all the nodes' vectors of the $i$th community as
$u_{1}^{i},u_{2}^{i},\cdots,u_{n_{i}}^{i}$. Let
$\bar{x}_{k}^{i}=\frac{1}{n_{i}}\sum_{j=1}^{n_{i}}u_{j}^{i}(k)\,
k=1,2\cdots,n$ be the mean characters of $i$th community,
$\bar{u}^{i}=(\bar{x}_{1}^{i},\bar{x}_{2}^{i},\cdots,\bar{x}_{n}^{i})$
be the $i$th community's barycenter and
$\bar{u}=(\bar{x}_{1},\bar{x}_{2},\cdots,\bar{x}_{n})$be all the
nodes' barycenter, here
$\bar{x}_{k}=\frac{1}{n}\sum_{j=1}^{n}x_{jk}\ (k=1,2,\cdots,n)$.
Then $F$ statistic is defined as
\begin{equation}
F=\frac{\sum_{i=1}^{c}\frac{n_{i}\parallel\bar{u}^{i}-\bar{u}\parallel^{2}}{c-1}}{\sum_{i=1}^{c}\sum_{j=1}^{n_{i}}\frac{\parallel
u_{j}^{i}-\bar{u}^{i}\parallel^{2}}{n-c}}, \label{F}
\end{equation}
 where
$\parallel\bar{u}^{i}-\bar{u}\parallel=\sqrt{\sum_{k=1}^{m}(\bar{x}_{k}^{i}-\bar{x}_{k})^{2}}$
is the distance between $\bar{u}^{i}$and $\bar{u}$, and $\parallel
u_{j}^{i}-\bar{u}^{i}\parallel$ is the distance between $u_{j}^{i}$
node of $i$th and the barycenter $\bar{u}^{i}$. The numerator of $F$
signifies the distance of inter-communities and the denominator the
distance of intra-communities. So the $F$ is bigger when the
difference distance of inter-communities is bigger and the
difference intra-communities is smaller. We can image that when $F$
achieve the maximum we can get the best partition.

\begin{figure} \center
\includegraphics[width=9cm]{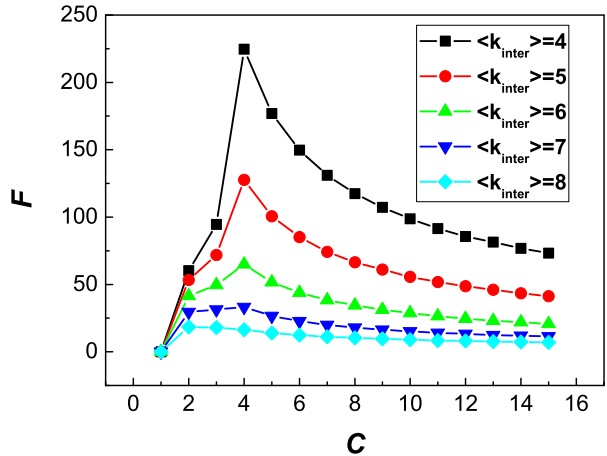}
\caption{$F$ statistic as the function of number of clusters $C$.
Plot shows the changes of $F$ statistic when $T=3$ and $\langle
k_{inter}\rangle$ changes from $4$ to $8$. It shows that when
$\langle k_{inter}\rangle$ is smaller than $8$, $F$ statistic can
identify the right number of communities. When the community
structure is clearer, the maximal value of $F$ statistic is very
distinct.}\label{Fpic}
\end{figure}

We use binary \emph{ad hoc} networks which contains $128$ nodes and
$4$ groups and proceed the signalling process as above to test the
$F$ statistic. The results show that $F$ statistic is very
efficacious. On the weighted \emph{ad hoc} networks, the results are
similar with binary ones. When the community structure is clearer
the maximal value of $F$ statistic is more distinct. The detailed
results are shown in Fig.\ref{Fpic}.

\subsection{Some Related Problems}

\begin{figure} \center
\includegraphics[width=9cm]{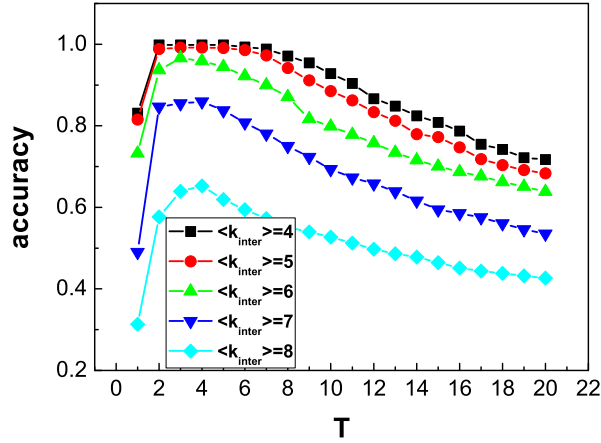}
\caption{ Let the number of communities be $4$ always. The plot
shows the changes of accuracy when $T$ changes form $1$ to $10$ on
binary \emph{ad hoc} networks. From the plots we can see $T\approx3$
is the proper length. And here $3$ is near the average shortest
paths of each artificial networks.}\label{Opt-T}
\end{figure}

\textbf{A. The Most Optimal $T$.} Parameter $T$ is an important
factor for the results of community identification. We can image
that the length $T$ must be sufficiently long to gather enough
information about the topology of the network and it should not be
too long to faint the information we have gathered. In order to let
majority of nodes can affect the whole network and do not to faint
the information about the topology of the network, we guess that it
may be optimal when $T$ is near to the average shortest path of
network. In order to demonstrate our guess, some numerical
experiments are done on binary networks which contains $128$ nodes
and $4$ groups as above. The results are shown Fig.\ref{Opt-T}. The
accuracy of the algorithm reach optimum when $T$ is 3 or 4. Of
course, we only do some numerical experiments, it is hard to say the
result satisfies all the networks. The random walks method\cite{6}
has also the same problem. How to find the most optimal $T$? We
think it is still an open problem.

\textbf{B. Time Complexity analysis.} The time complexity of
computing $F$ statistic is $O(n^2)$. For a definite $C$ which is the
number of communities, the time complexity for Fuzzy $C$-mean
clustering is  $O(Cn^2)$. Time complexity of the process of signal
diffusion is $O(Tn^3)$ when we use the multiplication of matrix to
simulate the process. But if we simulate the process in a network
directly, the corresponding time complexity is $O(T(k+1)n^2)$,
where, $k$ is the average degree of node in a network.

\textbf{C. Generalization to weighted networks.} It is easy to
generalize our algorithm to weighted network. Suppose we have a
weighted network with $n$ nodes, it can be represented
mathematically by an adjacency matrix $\textbf{W}$ with elements
$W_{ij}$. $W_{ij}$ denotes the connection strength of node $i$ and
$j$ (In some weighted network $W_{ij}$ dose not denote the strength
of connection, we should transform the weight before the algorithm),
then $\textbf{V}=(\textbf{I}+\textbf{W})^{T}$. The rest of the
algorithm is same with the algorithm on binary \emph{ad hoc}
network.

\textbf{D. Relations with other methods.} There two main differences
between our method and random walks method and circuit approach
method. First, we use the signal diffusion process to transfer the
topology to geometrical structure. The mathematical form is
$(\textbf{I}+\textbf{A})^{T}$. The distance of each pair of column
vectors of the matrix is the intimacy of the corresponding pair of
nodes. The random walk method gets the intimacy of each pair of
nodes by random walks. The mathematical form is
$(diag(\frac{1}{d_{1}}\frac{1}{d_{2}}\cdots\frac{1}{d_{n}})\textbf{A})^{T}$
where, $d_{i}$ denote the degree of node $i$, $diag$ means the
diagonal matrix. Take account of the effect of node degree, it also
use the Euclidian distance to define the intimacy. The circuit
approach method gets the intimacy of each pair of nodes by
Kirchhoff's law. Adding pressure on the proper two nodes, by
Kirchhoff's law, we can get the pressure of each node. More close of
the pressure of two nodes are, more intimate the two nodes are.
Suppose add pressure on nodes $1$ and $2$, then $p_{1}=1,p_{2}=0$,
where $p_{i}$ denotes the pressure of node $i$. The mathematical
form of Kirchhoff's law is
$\textbf{B}=(\textbf{A}diag(\frac{1}{d_{1}}\frac{1}{d_{2}}\cdots\frac{1}{d_{n}}))$,
$(p_{3},p_{4},\cdots,p_{n})^{'}=(I-\textbf{\~{B}})^{-1}C$, where
$\textbf{\~{B}}$ denotes the matrix $\textbf{B}$ with deleting the
first and the second columns and rows,
$C=(\frac{A_{31}}{d_{3}},\frac{A_{41}}{d_{4}},\cdots,\frac{A_{n1}}{d_{n}})^{'}$.
Second, as to the method of clustering, we use the $F$ statistic and
classical FCM method to partition the vectors. When the $F$
statistic achieve its maximum, we get the best partition. The random
walks method and the circuit approach method are all need the help
of other indexes to get the best partitions. One is modularity $Q$,
the other is tolerance. So we could say that $F$ statistic and FCM
method are all based on the geometrical structure of the vector
space, but the other two methods need the help of extra parameters.
Because the random walks method and the circuit approach method both
need some extra parameters which are very important to the results,
so in this paper we will not compare our results with that of these
two algorithms in different networks. Instead, we will compare the
accuracy and precision with other famous algorithms which do not
require any extra parameters.

\section {Results and Comparison with other Algorithms}

In order to investigate our algorithm, the accuracy and precision of
our algorithm will be compared with Potts algorithm(Potts)\cite{8},
Girvan-Newman algorithm(GN)\cite{22} and extremal optimal
algorithm(EO)\cite{5}. All these algorithms can be generalized to
weighted networks\cite{21}. Here we abbreviate GN weighted version
as WGN and EO as WEO. The accuracy and precision are defined in
\cite{21}. Accuracy means the consistence when the community
structure from algorithm is compared with the presumed communities,
and precision is the consistence among the community structures from
different runs of an algorithm on the same network. The algorithm's
accuracy and precision are calculated by $S$ function\cite{21} in
this paper. In the following numerical investigations on \emph{ad
hoc} networks, we first get $20$ realizations of artificial
community networks under the same conditions. Then we run each
algorithm to find communities in each network $10$ times. Based on
these results, using the similarity function $S$, comparing each
pair of these $10$ community structures and averaging over the $20$
networks (average of totally $C_{2}^{10}\times20=900$ results) we
could get the precision of the algorithm. Comparing each divided
groups with the presumed structures, we can get the accuracy of the
algorithm by averaging these $10\times20=200$ results.

\subsection {Results on \emph{ad hoc} networks}

\textbf{A. Binary \emph{ad hoc} networks.} In order to compare our
algorithm with others we first test it on computer-generated random
graphs with a well-known predetermined community structure\cite{22}.
Each graph has $N=128$ nodes divided into $4$ communities of $32$
nodes each. Edges between two nodes are introduced with different
probabilities depending on whether the two nodes belong to the same
group or not: every node has $\langle k_{intra}\rangle$ links on
average to its fellows in the same community, and $\langle
k_{inter}\rangle$ links to the outer-world, keeping $\langle
k_{intra}\rangle+\langle k_{inter}\rangle=16$. Fig.\ref{compare on
binary network} shows the results. The precision of our algorithm is
better than EO and almost the same as Potts and GN. While the
accuracy of our algorithm is better than GN and almost the same as
EO and Potts.

\begin{figure} \center
\includegraphics[width=7cm]{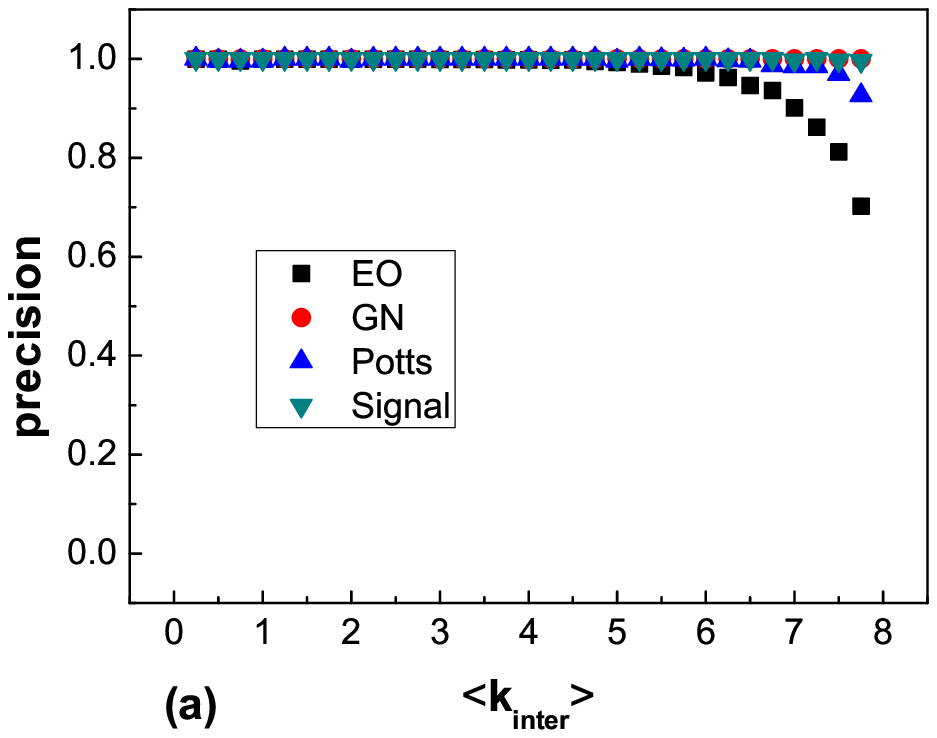}\includegraphics[width=7cm]{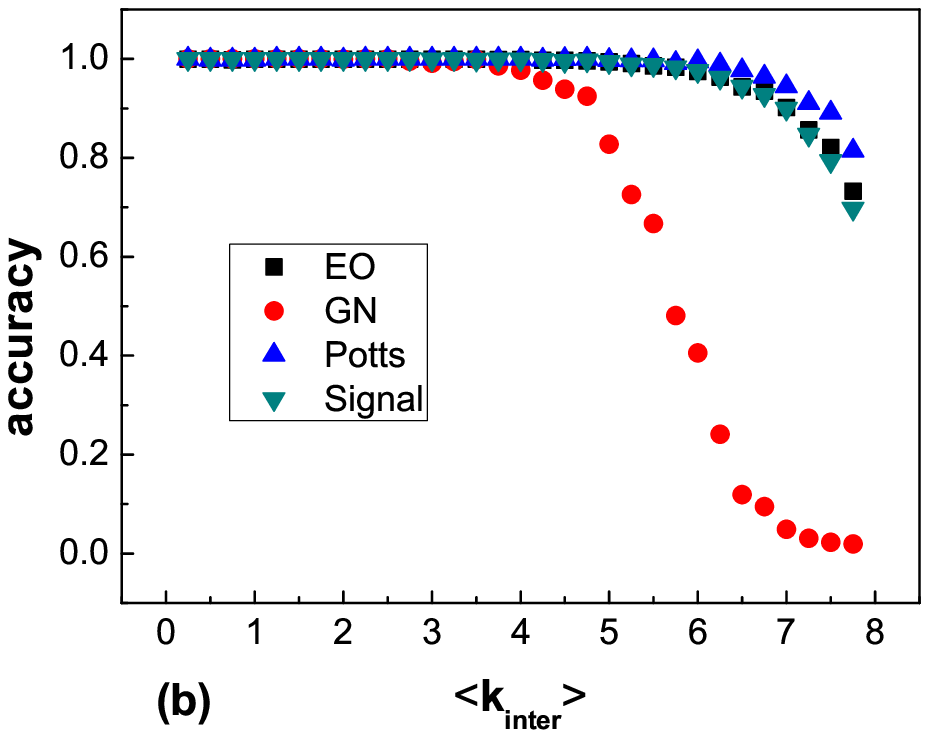}
\caption{Algorithm performance as applied to \emph{ad hoc} networks
with $n=128$ and four communities of $32$ nodes each. Total average
degree is fixed to $16$. We can see that the precision of our
algorithm is better than EO and Potts and equal GN and the accuracy
is better than GN and almost the same as EO and
Potts.}\label{compare on binary network}
\end{figure}

\begin{figure} \center
\includegraphics[width=7cm]{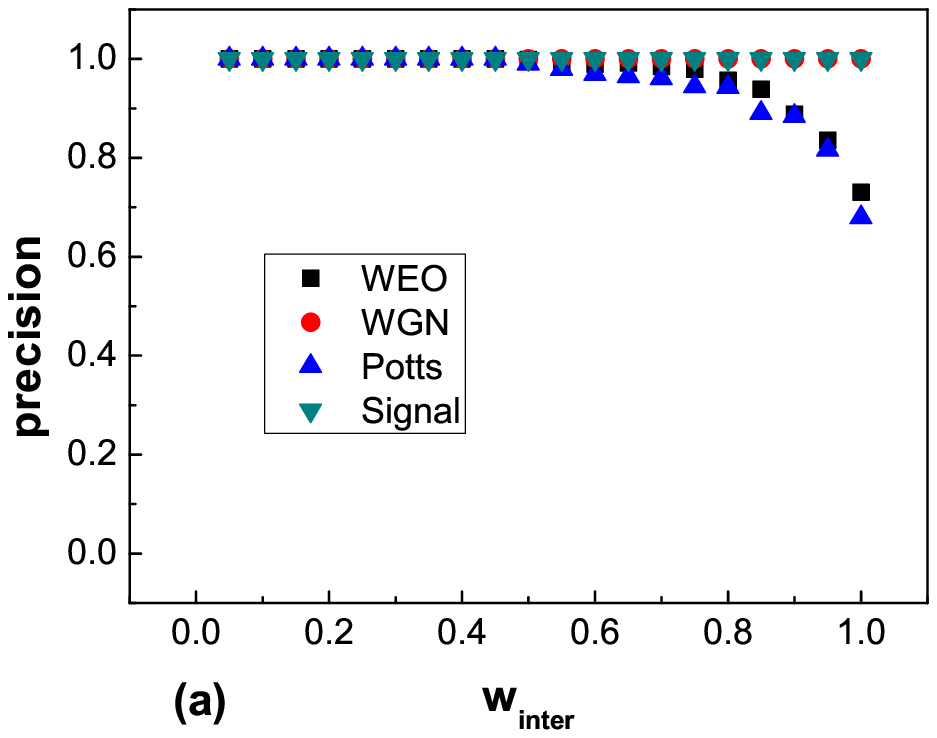}\includegraphics[width=7cm]{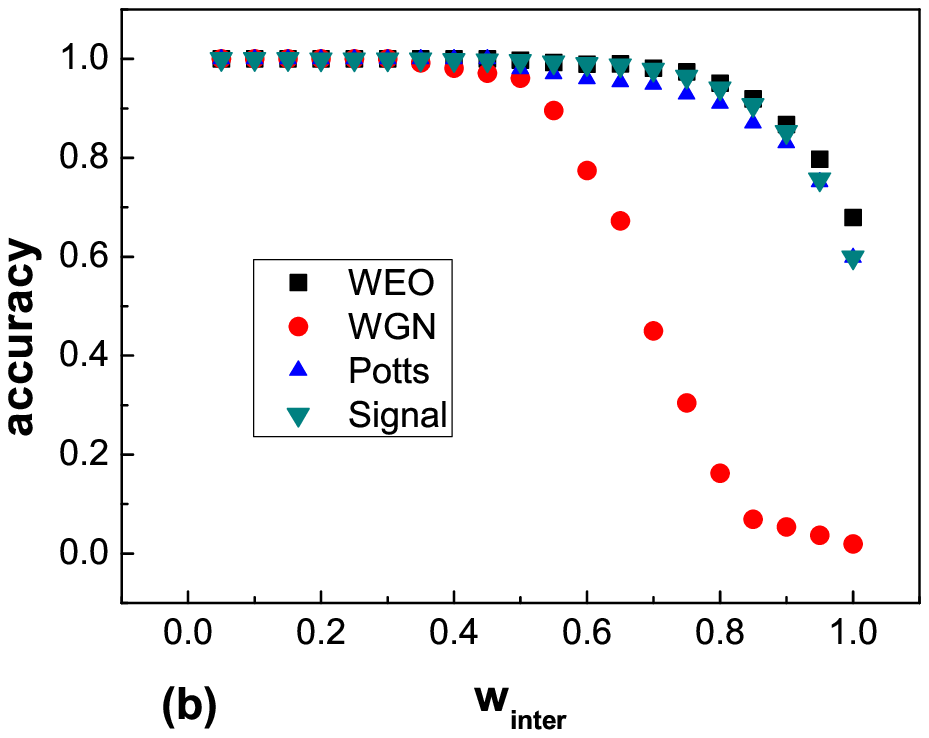}
\caption{The influence of weight on community structure when
$\langle k_{intra}\rangle=\langle k_{inter}\rangle=8$, where
$w_{inter}$ changes from $0.05$ to $1$. We can see the precision of
our algorithm is better than WEO and Potts and equal to WGN and the
accuracy of our algorithm is better than WGN but almost equals WEO
and Potts.}\label{compare wighted network 1}
\end{figure}

\textbf{B. Weighted \emph{ad hoc} networks. }In weighted networks,
we use similarity link weight to describe the closeness of relations
between nodes. Under the basic construction of \emph{ad hoc} network
described above, the intragroup link weight is assigned as
$w_{intra}$, while the intergroup link weight is assigned as
$w_{inter}$. Similarly with $\langle k_{intra}\rangle+\langle
k_{inter}\rangle=16$, we require the link weight on intra and inter
links follow the constraint:  $w_{intra}+ w_{inter}=2$, where
$w_{intra}$ ( $w_{inter}$) is the average of all intragroup
(intergroup) link weights. Here for simplicity, we assign the same
weight $w_{inter}=w$  to all intergroup links, and assign the same
weight $w_{intra}=2-w$ to all intragroup links. From
Fig.\ref{compare wighted network 1}, we can find that the precision
of our algorithm is better than WEO and Potts and equal to WGN, the
accuracy of our algorithm is better than WGN but almost equals WEO
and Potts. Even for the case with $\langle k_{inter}\rangle<\langle
k_{intra}\rangle$ but $\langle w_{inter}\rangle>\langle
w_{intra}\rangle$, or with uniform distribution of link weights, we
can get similar conclusions.

\begin{figure} \center
\includegraphics[width=7cm]{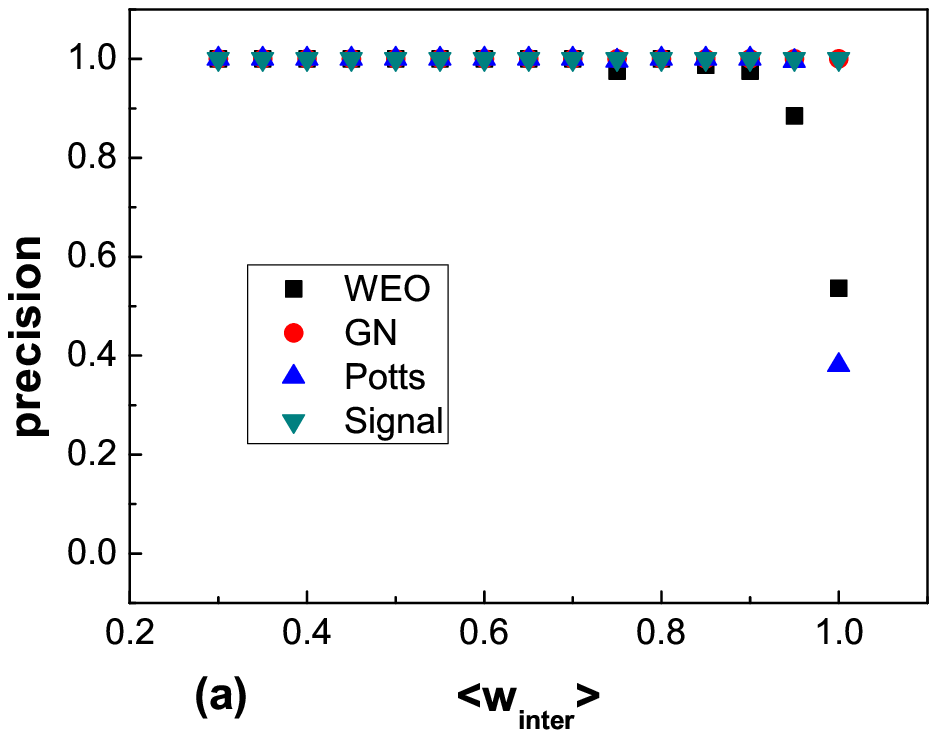}\includegraphics[width=7cm]{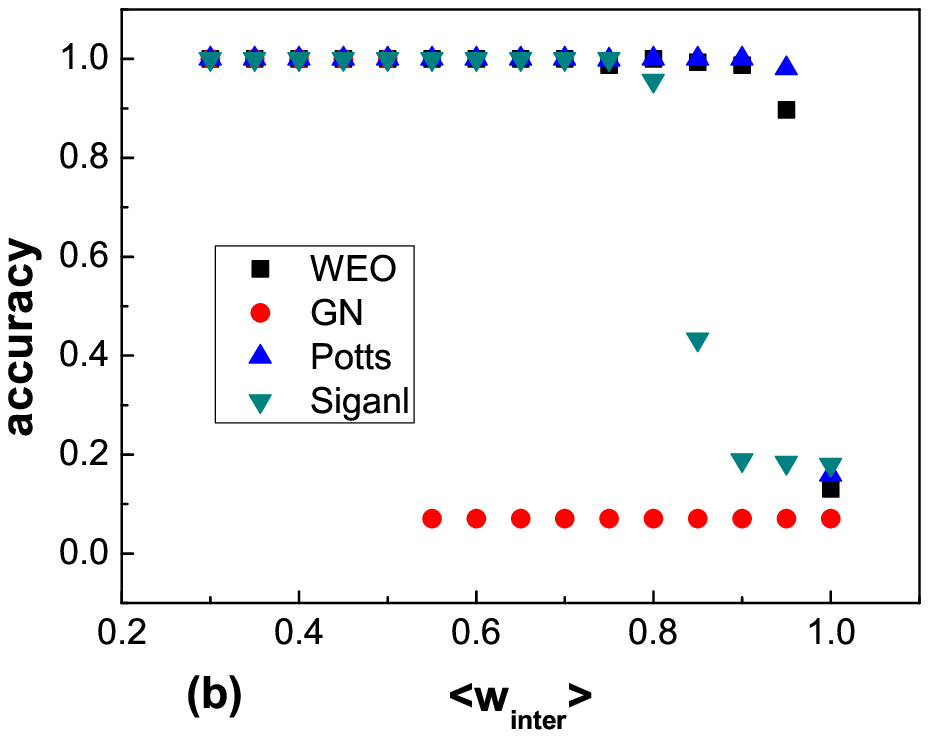}
\caption{Precision and accuracy of our algorithm and WGN, Potts, WEO
algorithms in complete networks with presumed communities. When
$\langle w_{inter}\rangle\gg\langle w_{intra}\rangle$, the precision
of our algorithm is better than WEO and Potts and equal to GN and
the accuracy of our algorithm is better than GN always but sharply
declines to near $0.2$ when  $\langle w_{inter}\rangle$ is greater
than $0.9$.}\label{compare on complete wighted networks}
\end{figure}

\textbf{C. Complete weighted networks. }An extreme idealized example
is the complete network. In complete networks, we use uniform
distribution of link weights. Weights are taken randomly from the
interval $[\langle w_{intra}\rangle-0.25, \langle
w_{intra}\rangle+0.25]$ and $[\langle w_{inter}\rangle-0.25, \langle
w_{inter}\rangle+0.25]$ respectively, for intragroup connections and
intergroup connections. The precision of our algorithm is better
than WEO and Potts and equal to WGN when $\langle
w_{inter}\rangle\ll\langle w_{intra}\rangle$, but its accuracy
almost declines to zero when $\langle w_{inter}\rangle$ is greater
than $0.9$. Fig.\ref{compare on complete wighted networks} shows the
results in detail.

\subsection{Empirical Results on Some Real Networks}

\textbf{A. Zachary's karate club.} Zachary karate club
network\cite{ZK} has bee considered as a simple sample for community
detecting methodologies\cite{2,11,13,15,19,22}. This network was
constructed with the data collected observing $34$ members of a
karate club over a period of $2$ years and considering friendship
between members. We let $T=3$ and get the best partition which
perfectly corresponds to the actual division of the club Fig.\ref{ZK
networks}

\begin{figure} \center
\includegraphics[width=8cm]{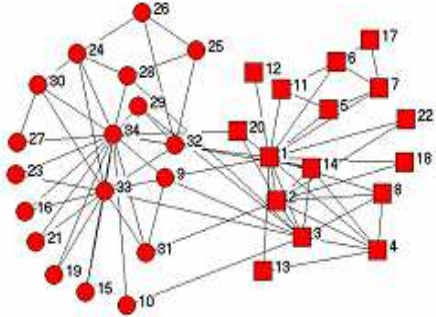}
\caption{Our algorithm detects $2$ communities form Zachary karate
club, which perfectly corresponds the actual division.}\label{ZK
networks}
\end{figure}

\begin{figure} \center
\includegraphics[width=8cm]{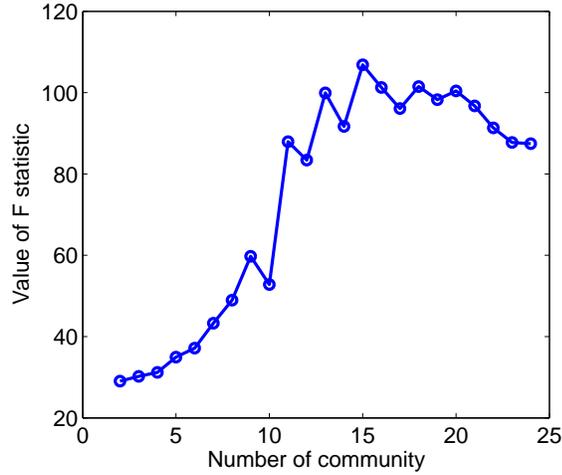}
\caption{Plot shows the change of the $F$ statistic with the number
of communities when $T=2$. $F$ statistic achieve its maximum when
the number of communities is $15$, among which $2$ communities just
have $2$ and $3$ teams respectively}\label{Footabll F}
\end{figure}

\textbf{B. College football network.} The algorithm is also applied
to College football network which was provided by Mark Newman. The
network is a representation of the schedule of Division I games for
the $2000$ season: vertices in the network represent teams and edges
represent regular-season games between the two teams they connect.
What makes this network interesting is that it incorporates a known
community structure. The teams are divided into $12$ conferences.
Games are more frequent between members of the same conference than
between members of different conferences. The average shortest path
length of the football network is $2.5$, so we let the signaling
time $T=3$. When the $F$ statistic achieve the maximum we get the
best partition. We also use the accuracy index of detecting
community algorithm \cite{21} to measure the effect of our algorithm
and find that it identifies the conference structure with a high
degree of success. We detect $15$ communities when $F$ reach it's
maximum (Fig.\ref{Footabll F}) among which $2$ communities just have
$2$ and $3$ teams respectively and five communities were detected
absolutely. The average accuracy is $0.78$ which is little better
than any of others (Tab.\ref{Table football}).

\begin{table}
\centering{Table 1. The accuracy of each detected community compared
with the counterpart of real world community}\label{Table football}
\\\begin{tabular} [t]{|c|c|c|c|c|}
\hline
% after \\: \hline or \cline{col1-col2} \cline{col3-col4} ...
Conference name&FCM accuracy&GN accuracy&EO accuracy&Potts
accuracy\\\hline Atlantic Coast&0.4444&1&1&1\\\hline Big
East&0.8000&0.8889&0.8000&0.8000\\\hline Big10&1&1&1&1\\\hline
Big12&1&0.9231&1&1\\\hline Conference
USA&0.9000&0.9000&0.9000&0.9000\\\hline IA
Independents&0.2500&0&0&0\\\hline Mid
American&0.7692&0.8667&0.9286&0.8667\\\hline Mountain
West&1&0&0.5714&0.5714\\\hline Pac10&1&0.5556&1&1\\\hline
SEC&1&0.7500&0.7059&0.7500\\\hline Sunbelt&0.4444&0.4444&0&0
\\\hline Western Athletic&0.7273&0.7273&0.7273&0.7273
\\\hline Average accuracy&0.7779&0.6713&0.7194&0.7179\\\hline
\end{tabular}
\end{table}

\section{Conclusion}
The investigation of community structures in complex networks is an
important issue in many domains and disciplines. This problem is
relevant to social tasks, biological inquiries, or technological
problems. In this paper, we have introduced a method to detect
communities based on the signaling process on networks.

In a complex networks with $n$ nodes, every node is viewed as a
system which can be inspired. Each node sends its neighbors and
itself signals and record the number of signals it receives at every
time step. For each node of the network, we give it an initial unit
quantity signal and other nodes have the signal of zero. Then after
the inspiration of $T$ steps on the network, the signal distribution
of the nodes denoted by an $n$-dimensional vector can be viewed as
the influence of the source node to the whole network. The amount of
signals be sent is equal to its present-owning signals. In complex
networks, we can generally consider that the node always influence
its community first then through the community influence the whole
network. So naturally, compared with the nodes of other communities,
nodes in the same community have similar influence toward the whole
network. So through the signaling process, the network partition
problem is transformed into the vectors clustering problem in
Euclidian space. The clustering can be work out by Fuzzy $C$-means
clustering method(FCM) with the help of $F$ statistic. Moreover, our
algorithm can also be generalized to weighted networks when we think
the weighted connections can magnify or dwindle the signals
linearly. Thus the method presented here can detect the optimal
community structures in binary and weighted networks without any
extra parameters.

To solve the partition problem of complex networks, precision and
accuracy of an algorithm are two standards for us to choose the
method. So we make a comparison between our algorithm and other
relatively mature ones such as EO, Potts and GN algorithms both in
binary and weighted networks. Results for both \emph{ad hoc} and
real networks have proved that our algorithm is effective. One
problem of our algorithm is that we haven't given a clear range of
the most optimal steps $T$ of inspiring. Actually, some other
algorithms such as the random walks method\cite{6} exist also the
similar problem. So we think it is an open problem and we will do
some deep research on it in the future.
\section*{Acknowledgement}
The authors thank Professor M.E.J. Newman very much for providing
College Football network data. This work is partially supported by
$985$ Projet and NSFC under the grant No.$70771011$, No.$70431002$,
and No.$60534080$.


\begin{thebibliography}{99}
\bibitem{1}R. Albert, A.-L. Barabasi, Rev. Mod. Phys. \textbf{74},
47 (2002).
\bibitem{Newman reviwe} M. E. J. Newman, SIAM Rev. \textbf{45},
167-256 (2003).
\bibitem{new reviwe} S. Boccaletti, V.Latora, Y. Moreno, M. Chavez, and D.-U. Hwang, Physics Report. \textbf{424},
175-308 (2006).
\bibitem{www} R. Albert, H. Jeong, A.-L. Barabasi, Nature.
\textbf{401},130 (1999).
\bibitem{social network1}S.Redner, Eur.Phys. J. B \textbf{4}, 131
(1998).
\bibitem{social network2}M. E. J. Newman, Proc. Natl. Acad. Sci, U. S. A.
\textbf{98},404 (2001).
\bibitem{metabolism}H. Jeong, B. Tombor, R. Albert, Z.N.Oltvai and A.-L.Barabasi, Nature
\textbf{407},651 (2000).
\bibitem{14}M. E. J. Newman,  Proc. Natl. Acad. Sci. U. S. A \textbf{103},
8577-8582 (2006).
\bibitem{15}M. Girvan and M. E. J. Newman, Proc. Natl. Acad. Sci. U. S. A
\textbf{99}, 7821-7826 (2004).
\bibitem{2}M. E. G Newman, Phys. Rev. E \textbf{69}, 066133
(2004).
\bibitem{3}L. Danon, J. Duch, A. Arenas, and A. Diaz-Guilera, J. Stat. Mech. P09008
(2005).
\bibitem{4}S. Lehmann and L. K. Hansen, arxiv.org/abs/physics/0701348
(2007).
\bibitem{5}J. Duch and A. Arenas, Phys. Rev. E \textbf{72}, 027104
(2005).
\bibitem{6}M. Latapy and P. Pons, Computing communities in large networks using random walks, in Proceedings of the 20th International Symposium on Computer and Information Sciences, ISCIS'05, LNCS 3733, 284-293
(2005).
\bibitem{7}F. Radicchi, C.o Castellano, F. Cecconi, V. Loreto, and D. Parisi, Proc. Natl. Acad. Sci. U.S.A \textbf{101},
2658 (2004).
\bibitem{8}J. Reichardt and S. Bornholdt, Phys. Rev. Lett. \textbf{93},
218701 (2004).
\bibitem{9}F. Wu and B. A. Huberman, The Eur. Phys. J. B
\textbf{38},331-338 (2004).
\bibitem{10}A. Clauset, Phys. Rev. E \textbf{72}, 026132
(2005).
\bibitem{11}J. P. Bagrow and E. M. Bollt, Phys. Rev. E \textbf{72},
046108 (2005).
\bibitem{12}S. Muff, F. Rao and A. Caflisch, Phys. Rev. E \textbf{72},
056107 (2005).
\bibitem{13}M. E. J. Newman and E. A. Leicht, Proc. Natl. Acad. Sci. USA
104, 9564-9569 (2007).
%\bibitem{16}B. S. Everitt, S. Landau, and M. Leese, Cluster Analysis, Hodder Arnold, London, 4th edition,
%(2001).
\bibitem{18}C. P. Massen and J. P. K. Doye, Phys. Rev. E \textbf{71},
046101 (2005).
\bibitem{19}L. Donetti and M. A. Munoz, J. Stat. Mech. P10012
(2004).
\bibitem{20}A. Capocci, V. D. P. Servedio, G. Caldarelli, and F. Colaiori, Physica A \textbf{352},
669 (2005).
\bibitem{22}M. E. J. Newman, M. Girvan, Phys. Rev. E \textbf{69},
026113 (2004).
\bibitem{Q NPC}U. Brandes, D. Delling, M. Gaertler, R. Gorke, M. Hoefer, Z. Nikoloski, and D. Wagner, arXiv:physics/0608255,
(2006).
\bibitem{23}S. Fortunato and M. Barthelemy, Natl. Acad. Sci. U. S. A. 104, 36
(2007).
\bibitem{ZK}W. Zachary, Journal of Anthropol Research, \textbf{33}, 452
(1977).
\bibitem{17}M. E. J. Newman,  Phys. Rev. E \textbf{74}, 036104
(2006).
\bibitem{21}Y. Fan, M. Li, P. Zhang, J. Wu, Z. Di, Physica A 377
(2007).
\bibitem{24}A. Li, Fuzzy mathematics and application. Metallurgical Industry Press. Beijing, (2005). (Chinese book.ISBN-7-5024-3818-1).
\bibitem{Actual ZK}Image courtesy Mark Newman's site at http://www.personal.umich.edu/mejn/networks/

\end{thebibliography}
\end{document}